\documentclass[conference]{IEEEtran}
\IEEEoverridecommandlockouts

\usepackage{cite}
\usepackage{xcolor} 

\usepackage{units}
\usepackage{mathtools} 
\usepackage{amsthm}
\newcounter{thm}
\newtheorem{prob}[thm]{Problem}

\usepackage{pgfplots}
\usepgfplotslibrary{groupplots}

\usepackage{hyperref}
\usepackage{amsmath}
\usepackage{amsfonts}
\usepackage{graphicx} 
\usepackage{amssymb}  
\usepackage{pstricks,pst-plot,psfrag}
\usepackage{float}
\usepackage{dsfont}

\usepackage{lettrine}
\usepackage{import}
\usepackage[acronym]{glossaries}

\usepackage{tikz}
\usetikzlibrary{backgrounds,calc,angles,quotes,arrows.meta,arrows,fit,shapes.geometric,shapes.multipart,positioning}
\usetikzlibrary{circuits.ee.IEC}
\usetikzlibrary{shapes.gates.logic.US}

\newacronym{acr:cvt}{CVT}{continuously variable transmission}
\newacronym{acr:CoG}{CoG}{center of gravity}
\newacronym{acr:CoP}{CoP}{center of pressure}
\newacronym{acr:CV}{CV}{constant-velocity}
\newacronym{acr:dp}{DP}{dynamic programming}
\newacronym{acr:DoF}{DoF}{degrees of freedom}
\newacronym{acr:ecms}{ECMS}{equivalent consumption minimization strategies}
\newacronym{acr:eltms}{ELTMS}{equivalent lap time minimization strategies}
\newacronym{acr:em}{EM}{electric motor}
\newacronym{acr:es2k}{ES2K}{Energy Storage to Kinetic}
\newacronym{acr:F1}{F1}{Formula 1}
\newacronym{acr:FIA}{FIA}{F\'{e}d\'{e}ration Internationale de l'Automobile}
\newacronym{acr:fgt}{FGT}{fixed-gear transmission}
\newacronym{acr:FD}{FD}{final drive}
\newacronym{acr:ice}{ICE}{internal combustion engine}
\newacronym{acr:k2es}{K2ES}{Kinetic to Energy Storage}
\newacronym{acr:mgu}{MGU}{motor generator unit}
\newacronym{acr:mguh}{MGU-H}{motor generator unit heat}
\newacronym{acr:mguk}{MGU-K}{motor generator unit kinetic}
\newacronym{acr:mpc}{MPC}{model predictive control}
\newacronym[description={energy management strategy}, \glslongpluralkey={energy management strategies},\glsshortpluralkey={EMSs}]{EMS}{EMS}{energy management strategy}%
\newacronym{acr:ODE}{ODE}{ordinary differential equation}
\newacronym{acr:pmp}{PMP}{Pontryagin's Minimum Principle}
\newacronym{acr:pu}{PU}{power unit}
\newacronym[description={powertrain operation}, \glslongpluralkey={powertrain operations},\glsshortpluralkey={POs}]{acr:PO}{PO}{powertrain operation}%
\newacronym{acr:socp}{SOCP}{second-order cone program}
\newacronym{acr:soe}{SoE}{state of energy}

\newcommand{\pushright}[1]{\ifmeasuring@#1\else\omit\hfill$\displaystyle#1$\fi\ignorespaces}
\newcommand{\pushleft}[1]{\ifmeasuring@#1\else\omit$\displaystyle#1$\hfill\fi\ignorespaces}
\makeatother
\newif\ifmargincomments 
\margincommentstrue

\newif\ifextendedversion 
\extendedversiontrue

\ifmargincomments

\newcommand{\jkmargin}[2]{{\color{blue}#1}\marginpar{\color{blue}\raggedright\footnotesize [JK]:#2}}

\else

\newcommand{\jkmargin}[2]{#1}
\fi
\usetikzlibrary{external}
\begin{document}

\title{A Two-dimensional Spatial Optimization Framework for Vehicle Powertrain Systems\\
\thanks{Jorn van Kampen, Mauro Salazar and Theo Hofman are with the Control Systems Technology section, Department of Mechanical Engineering, Eindhoven University of Technology (TU/e), Eindhoven, 5600 MB, The Netherlands.
E-mails: {\tt\footnotesize j.h.e.v.kampen@tue.nl}, {\tt\footnotesize m.r.u.salazar@tue.nl}, {\tt\footnotesize t.hofman@tue.nl} \newline
This paper was supported by the NEON research
project (project number 17628 of the Crossover program which
is (partly) financed by the Dutch Research Council (NWO)). }
}

\author{\IEEEauthorblockN{Jorn van Kampen, Mauro Salazar and Theo Hofman}
%

}

\maketitle

\begin{abstract}
This paper presents a modeling framework to optimize the two-dimensional placement of powertrain elements inside the vehicle, explicitly accounting for the rotation, relative placement and alignment. 
Specifically, we first capture the multi-level nature of the system mathematically, and construct a model that captures different powertrain component orientations. %
Second, we include the relative element placement as variables in the model and derive alignment constraints for both child components and parent subsystems to automatically connect mechanical ports.
Finally, we showcase our framework on a four-wheel driven electric vehicle. Our results demonstrate that our framework is capable of efficiently generating system design solutions in a fully automated manner, only using basic component properties. 
%
\end{abstract}


\section{Introduction}\label{sec:introduction}
The design of mechatronic systems, such as electric vehicle powertrains, is challenging due to their sheer complexity. Key performance indicators, such as the overall powertrain efficiency and cost, are influenced by both the components and the system architecture, resulting in the need of a holistic approach. Such a holistic approach typically aims at finding the optimal topology, being the one-dimensional element connections, through computational design synthesis methods~\cite{WijknietHofman2018}. However, once a suitable topology is found, there exists an almost infinite amount of possible physical implementations. Moreover, the physical placement of the components once again impacts the overall performance. In the case of a powertrain, the component placement directly affects the vehicle weight distribution, length of connections and physical spacing which, in turn, affects the total vehicle weight and thereby the energy consumption. Furthermore, a more compact placement of the components can increase passenger space within the vehicle, thereby improving comfort. Therefore, it is important to consider the spatial system design within the holistic design approach. Against this backdrop, this paper presents an optimization framework to compute the optimal powertrain element placement from a systems level perspective.

\begin{figure}[!t]
	\centering
	\tikzstyle{Component} = [rectangle, minimum width=0.2cm, minimum height=0.2cm,text centered, draw=black, line width=0.25mm]

\tikzstyle{simpleNodeWide} = [rectangle, minimum width=1.5cm, minimum height=1cm,text centered, draw=black, line width=0.25mm]
\tikzstyle{arrow} = [->,-triangle 45,line width=0.25mm,black]
\tikzstyle{lineE} = [-,line width=0.5mm,gray!50,solid]
\tikzstyle{lineM} = [-,line width=0.25mm,black]

\definecolor{HVorange}{HTML}{FF6700}%
\definecolor{colEM}{HTML}{C81919}%
\definecolor{colInv}{rgb}{0.5176,0.8235,0}%
\definecolor{colDiff}{HTML}{5858e8}%


\tikzset{
	relative at/.style n args = {3}{
		at = {({$(#1.west)!#2!(#1.east)$} |- {$(#1.south)!#3!(#1.north)$})}
	}
}
\begin{tikzpicture}
	tikzstyle{every node}=[font=\footnotesize] 
	
	\pgfmathsetmacro{\xDist}{0.75}
	\coordinate (Zero) at (0,0);
	
	\node[Component,fill=HVorange] (bat1) at(Zero) {};
	\node[Component,fill=HVorange, above= 0.5 cm of bat1] (batN)  {};
	\draw[dotted,very thick] (bat1)--(batN);
	
	\node [fit=(bat1)(batN), draw, thick, inner sep=2pt,label={above:Battery pack},align=center,anchor=center] (Bat) {};
	
		\node[Component,fill=colInv, right= \xDist cm of Bat,label={above:INV2}] (Inv2)  {};
		\node[Component,fill=colInv, above= \xDist cm of Inv2,label={above:INV1}] (Inv1)  {};
		\node[Component,fill=colInv, below= \xDist cm of Inv2,label={above:INV3}] (Inv3)  {};
		
		\draw[very thick,color=HVorange] (Bat.east)--(Inv1.west);
		\draw[very thick,color=HVorange] (Bat.east)--(Inv2.west);
		\draw[very thick,color=HVorange] (Bat.east)--(Inv3.west);
		
		\node[Component,fill=colEM, right= \xDist cm of Inv1,label={above:EM1}] (EM1)  {};
		\node[Component,fill=colEM, right= \xDist cm of Inv2,label={above:EM2}] (EM2)  {};
		\node[Component,fill=colEM, right= \xDist cm of Inv3,label={above:EM3}] (EM3)  {};
		\draw[very thick,color=HVorange] (Inv1.east)--(EM1.west);
		\draw[very thick,color=HVorange] (Inv2.east)--(EM2.west);
		\draw[very thick,color=HVorange] (Inv3.east)--(EM3.west);

		\node[Component,fill=gray!40, right= \xDist cm of EM2] (GP1)  {};
		\node[Component,fill=gray!40, right= \xDist cm of EM3] (GP2)  {};

		\draw[very thick,color=black] (EM2.east)--(GP1.west);
		\draw[very thick,color=black] (EM3.east)--(GP2.west);
		
		\node[Component,fill=gray!40, right= 0.2 cm of GP1] (GP11)  {};
		\node[Component,fill=gray!40, right= 0.2 cm of GP2] (GP22)  {};
		\draw[very thick,color=black] (GP1.east)--(GP11.west);
		\draw[very thick,color=black] (GP2.east)--(GP22.west);
		
		\node [fit=(GP1)(GP11), draw, thick, inner sep=2pt,label={above:TM1},align=center,anchor=center] (TM1) {};
		\node [fit=(GP2)(GP22), draw, thick, inner sep=2pt,label={above:TM2},align=center,anchor=center] (TM2) {};
		
		\node[Component,fill=colDiff, label={above:Diff}]  (Diff) at ($(TM1|-EM1)$)  {};
		\draw[very thick,color=black] (EM1.east)--(Diff.west);
				
		\node[Component,fill=gray, right= \xDist cm of TM1,label={right:RL}] (RL)  {};
		\node[Component,fill=gray, right= \xDist cm of TM2,label={right:RR}] (RR)  {};
		\node[Component,fill=gray,label={right:FL}] (FL) at ($(RL|-Diff)$) {};
		\node[Component,fill=gray,label={right:FR}] (FR)  at ($(RL)!0.5!(FL)$) {};
		\draw[very thick,color=black] (Diff.east)--(FR.west);
		\draw[very thick,color=black] (Diff.east)--(FL.west);
		\draw[very thick,color=black] (GP11.east)--(RL.west);
		\draw[very thick,color=black] (GP22.east)--(RR.west);

		\node[ inner sep=0pt,fill=white,anchor=west] (Packaging) at ($(RL)!0.5!(RL)+(0:2cm)$) 
		{\resizebox{1.5cm}{!} {\input{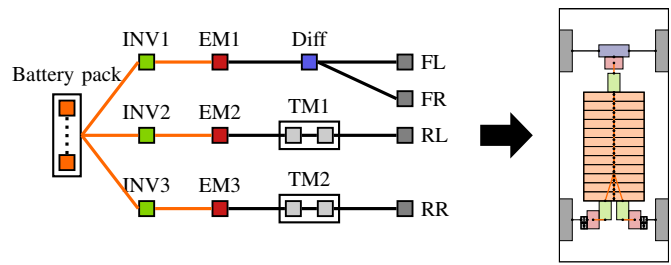}}};
		\draw[-{Triangle[length=3mm,angle'=90]},fill=black,minimum height=3em, line width=8pt] ($(RL)!0.5!(RL)+(0:1cm)$)--($(Packaging.west)+(180:0.3cm)$);
\end{tikzpicture}                
	\caption{Schematic representation of the workflow, whereby we translate a powertrain topology to a spatial implementation. The topology represents a four-wheel driven electric vehicle, where the front wheels are propelled by a single \gls{acr:em} through a differential, and the rear wheels are propelled by an individual \gls{acr:em} through a transmission. The battery pack and transmissions are both subsystems consisting of 24 modules and 2 gear pairs, respectively. The example presents a typical spatial design for the given topology.}
	\label{fig:topology}
\end{figure}

\subsubsection*{Related Literature}
This work pertains to two main research streams: design optimization for the main powertrain components and the topology, and (general) placement problems.

Several works have considered the optimization of the design and control of the powertrain through decomposed optimization~\cite{FahdzyanaSalazarEtAl2021,PourabdollahMurgovskiEtAl2014,SezerGokasanEtAl2011}. More recently, joint optimization of the design and control through nested~\cite{EbbesenDoenitzEtAl2012, ZhuChenEtAl2020} or simultaneous optimization~\cite{KondaHofmanEtAl2022, PourabdollahMurgovskiEtAl2013} has become increasingly popular, since it extends the design space. 
Several studies have further researched this field to also account for the powertrain topology optimization or generation to extend the search space~\cite{SilvasHofmanEtAl2015, WijknietHofman2018, KabalanVinotEtAl2020, ChenouardHartmannEtAl2016, KortWijknietEtAl2020}. The authors often apply a sequential or iterative approach in which a set of feasible topologies is generated first, after which the optimal solution is found using a drive cycle based optimization framework. However, these papers still only consider the topology and not the physical placement of elements in the vehicle. 

The second research stream involves placement problems, in which typically the bounding box of a set of elements is minimized through strategic placement~\cite{BoydVandenberghe2004}. These volumetric optimization problems are generally studied for small systems, such as transmissions~\cite{BerxGadeyneEtAl2014, PiacentiniCheongEtAl2020}, a washing machine~\cite{BerxFriedlEtAl2016} or a compressor box~\cite{RosichBerxEtAl2016}. While these studies consider both two- and three-dimensional problems, the problems are often simplified by fixing some degree of freedom, such as the element rotation or relative position between elements. 
A variation to volumetric optimization problems are (bin) packing problems, where all elements have to be placed in the minimal number of predefined spaces, as opposed to minimizing the bounding box. While these problems are NP-hard~\cite{LodiMartelloEtAl2002}, there exist several methods that approximate the solution through deterministic heuristics~\cite{ BansalLodiEtAl2005, Epstein2010, ElhedhliGzaraEtAl2019 } or stochastic algorithms~\cite{SzykmanCagan1997,AladahalliCaganEtAl2007}. Yet these types of problems do not capture constraints related to element connections.
 
In conclusion, to the best of the authors’ knowledge, there are ample opportunities to derive a method for optimizing the powertrain element placement from a systems perspective across multiple system levels, whilst accounting for alignment constraints.

\subsubsection*{Statement of Contributions}
This paper presents an optimization framework to automatically compute the two-dimensional placement and orientation of powertrain elements in the vehicle together with the optimal sizing of subsystems for an arbitrary topology. We approach the problem from a systems perspective and jointly optimize the spatial placement over multiple system levels and energy domains in the powertrain. 
To this purpose, we include the dimensions of subsystems containing internal elements as variables in the optimization problem. 
We frame the problem as a mixed-integer convex quadratic program, whereby the binary variables are used for the orientation and relative placement of the elements. 
Subsequently, we derive alignment constraints for mechanical connections as a function of the relative placement to automatically align mechanical connection ports. 
Finally, we showcase our framework for a four-wheel driven powertrain topology illustrated in Fig.~\ref{fig:topology}.

\subsubsection*{Organization}
The remainder of this paper is structured as follows: Section~\ref{sec:methods} presents the spatial optimization problem and highlights some of the most important aspects that are captured in our model. We demonstrate our framework in Section~\ref{Results} for a predefined topology. Finally, Section~\ref{Conclusion} draws the conclusions and provides an outlook on future research.

\section{Methods}\label{sec:methods}
This section presents the component placement problem for a two-dimensional mechatronic system consisting of an \textit{a priori} known set of connections and element properties, as illustrated by Fig.~\ref{fig:topology}. We apply a holistic approach by jointly optimizing across multiple system levels. Our model allows for various degrees of freedom of the elements, including the longitudinal and lateral position, rotation around the vertical axis, and flipping along the longitudinal and lateral centerlines. Depending on the energy domain (e.g., mechanical or electrical) of the element connection port, an element is constrained in terms of orientation and (relative) position to ensure a proper and feasible connection. This procedure is further elaborated in the next sections. First, we provide an overview of the modeling approach for the subsystems and components. Second, we derive constraints that allow different component orientations based on some binary decision variables. Third, we formulate interference constraints between element pairs to prevent infeasible solutions. Finally, we derive a method to automatically impose alignment constraints as a function of the relative placement between connected elements and formulate the resulting optimization problem. 

\subsection{System Modeling}







In this study, we regard two-dimensional mechatronic systems consisting of an arbitrary set of elements, where each element contains an arbitrary amount of internal system levels. For any system level $\lambda \in \{0,\ldots,N_\mathrm{levels}\} \subseteq \mathbb{N}$, we denote the set of elements by $V_\lambda \coloneqq \left \{ v_{\lambda,k} \ |\   k \in \mathcal{K}_\lambda \right \} = \mathcal{S}_\lambda \cup \mathcal{C}_\lambda $, where $ \mathcal{K}_\lambda$ is the set of identifiers for the elements at level $\lambda$, and $\mathcal{S}_\lambda$ and $\mathcal{C}_\lambda$ are the complete set of subsystems and components, respectively. Note that the set $v_{\lambda,k}$ represents element $k$ of level $\lambda$ that contains the elements belonging to the lower level of that element.
Here, we refer to a subsystem as an element that contains at least one lower system level with at least two elements present at this lower level, such as the battery pack in Fig.~\ref{fig:topology}. In contrast, we refer to a component as an element that does not contain any lower-level elements, such as the \gls{acr:em} in Fig.~\ref{fig:topology}. 
Elements can be connected to each other through their connection points, where the type of connection depends on the energy domain. Hereby, an element can have multiple connection points or \emph{ports} with an amount equal to the \emph{degree} of the element. We define the complete set of ports per element as $C_{\lambda,k} \coloneqq \left\{ c_{\lambda,k,i} | i \in \mathcal{I}_{\lambda,k} \right\} $, where $\mathcal{I}_{\lambda,k}$ is the set of identifiers for the ports of element $v_{\lambda,k}$. 
We denote the complete set of connections on a certain system level by $E_\lambda \coloneqq \{ (c_{\lambda,k,i},c_{\lambda,l,j}) | k, l \in \mathcal{K}_\lambda, \ i \in \mathcal{I}_{\lambda,k}, \ j \in \mathcal{I}_{\lambda,l} \}= \mathcal{M}_\lambda \cup \mathcal{E}_\lambda$, where $\mathcal{M}_\lambda$ represents the set of mechanical connections and $\mathcal{E}_\lambda$ represents the set of electrical connections. Similarly, we denote the set of connections between two system levels as $E_{\lambda,\lambda+1} \coloneqq \{ (c_{\lambda,k,i},c_{\lambda+1,l,j}) | k \in \mathcal{K}_\lambda, \ l \in \mathcal{K}_{\lambda+1},\  i \in \mathcal{I}_{\lambda,k}, \ j \in \mathcal{I}_{\lambda+1,l} \} = \mathcal{M}_{\lambda,\lambda+1} \cup \mathcal{E}_{\lambda,\lambda+1}$, with $\mathcal{M}_{\lambda,\lambda+1}$ being the set of mechanical connections and $\mathcal{E}_{\lambda,\lambda+1}$ being the set of electrical connections between the two system levels.


We automatically generate constraints based on several element properties that are obtained from the element library.
First, we require the component dimensions in terms of length $l$ and width $w$. Thereby, we model every component as a rectangle and in case of non-rectangular components, either include the dimensions of their bounding box or compose them of a set of rectangles. This allows us to simplify the representation of subsystems and interference constraints. In contrast to components, subsystems do not have fixed dimensions, since they are to be generated by optimally placing their internal elements within their bounding box defined by
\par\nobreak\vspace{-5pt}
\begingroup
\allowdisplaybreaks
\begin{small}
\begin{subequations}
\begin{alignat}{3}\label{eq:El_in_box}
	\frac{w_{\lambda,i}}{2}-\frac{w_{\lambda-1,j}}{2} &\leq x_{\lambda-1,j}-x_{\lambda,i} && \leq \frac{w_{\lambda-1,j}}{2}-\frac{w_{\lambda,i}}{2}, \\
	\frac{l_{\lambda,i}}{2}-\frac{l_{\lambda-1,j}}{2} &\leq y_{\lambda-1,j}-y_{\lambda,i} && \leq \frac{l_{\lambda-1,j}}{2}-\frac{l_{\lambda,i}}{2}, 
\end{alignat}
\end{subequations}
\end{small}%
\endgroup
for all elements in a subsystem, i.e., for all $v_{\lambda,i} \in v_{\lambda-1,j}$, for all $ j \in \mathcal{K}_{\lambda-1}$, for all $ \lambda\in  \{1,\ldots,N_\mathrm{levels}\}$. Here, $x_{\lambda,i}$ and $y_{\lambda,i}$ denote the lateral and longitudinal position within the design space, respectively. 
Second, we require connection properties per port for both components and subsystems. These properties include the energy domain of the port and connection type (direct or indirect). For components, the relative location of each connection port w.r.t. the component center is fixed, which means that this is required as an additional component property. In the case of subsystems, the location of the port is an optimization variable. For electrical ports, the location of the port is free within the bounding box of the element, given by
\par\nobreak\vspace{-5pt}
\begingroup
\allowdisplaybreaks
\begin{small}
\begin{subequations}
\begin{alignat}{3}
	-\frac{w_{\lambda,k}}{2} &\leq a_{\lambda,k,i} &&\leq \frac{w_{\lambda,k}}{2}, \label{eq:CP_in_box1}\\
	-\frac{l_{\lambda,k}}{2} &\leq b_{\lambda,k,i} &&\leq \frac{l_{\lambda,k}}{2}, \label{eq:CP_in_box2}
\end{alignat}
\end{subequations}
\end{small}%
\endgroup
where $(a_{\lambda,k,i}, b_{\lambda,k,i}) \in C_{\lambda,k}$ for all $ v_{\lambda,k} \in V_\lambda$, for all $ \lambda\in  \{0,\ldots,N_\mathrm{levels}\}$, represent the relative x- and y-location with respect to the center of the element. 
For mechanical ports, we specify the connection port to be on the edge of the component bounding box to represent a shaft connection. Since this is not a trivial constraint formulation, we will explain it in more detail in Section~\ref{sec:alignment}. 



\subsection{Component Orientation}
In the element library, we specify the component dimensions for a default orientation. However, we want to explore different orientations, which means that the component should be rotatable. To this effect, we limit the change in orientation of elements to steps of \unit[90]{degrees}.   
First, we introduce a binary variable $r_{\lambda,k}$ to rotate the component shape through
\par\nobreak\vspace{-5pt}
\begingroup
\allowdisplaybreaks
\begin{small}
\begin{subequations}
\begin{alignat}{4}\label{eq:rotate}
	w_{\lambda,k} &=  \overline{w}_{\lambda,k}&\cdot (1-r_{\lambda,k}) &+ \overline{l}_{\lambda,k}&\cdot r_{\lambda,k} , \\
	l_{\lambda,k} &=  \overline{l}_{\lambda,k}&\cdot (1-r_{\lambda,k}) &+ \overline{w}_{\lambda,k}&\cdot r_{\lambda,k} , 
\end{alignat} 
\end{subequations}
\end{small}%
\endgroup  
where the constraints hold for all components, i.e., for all $v_{\lambda,k} \in \mathcal{C}_\lambda$, for all $\lambda \in \{0,\ldots,N_\mathrm{levels}\}$. Note that a change of orientation is only possible for components and not for subsystems, since the dimensions of the latter are jointly optimized, thereby removing the need for different orientations.
Second, we add two more binary variables $m_{\lambda,k}$ and $n_{\lambda,k}$ to flip ports along the component's longitudinal and lateral centerline, respectively, through
\par
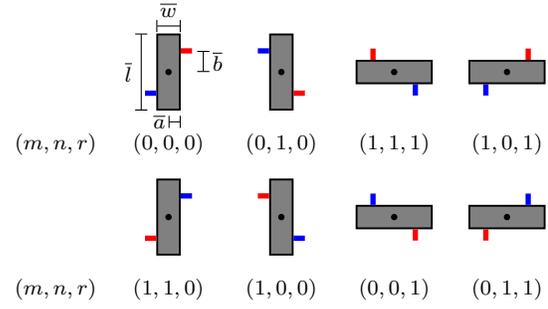
\begin{figure}[!t]
	\centering
	\tikzstyle{Comp} = [rectangle, minimum width=0.5cm, minimum height=1cm,text centered, draw=black, line width=0.25mm,fill=gray]

\tikzset{
	pics/CompVar/.style args = {#1,#2}{
		code ={
			\node[rectangle, minimum width=0.3cm, minimum height=1cm,text centered, draw=black, line width=0.25mm,fill=gray,rotate=#1,align=center] (-Comp) {};
		\draw[line width=0.7mm,red] (-Comp.#2) --++ (#1:0.15) node (-in) {};
		\draw[line width=0.7mm,blue] (-Comp.#2-180) --++ (#1:-0.15) node (-out) {};
		\filldraw[black] (-Comp.center) circle (1pt); 
	}
	}
}

\begin{tikzpicture}

	\tikzstyle{every node}=[font=\footnotesize] 
	
	\definecolor{Fcol}{rgb}{1,0.00,0.00} 
	\definecolor{vcol}{rgb}{0,0.50,0.00} 
	
	\coordinate (Origin) at (0,0);
	\pgfmathsetmacro{\xDist}{1.5}
	\pgfmathsetmacro{\yDist}{0.7}

	\pic[inner sep=0pt,anchor=center] (n1) at (0,0)  {CompVar={0,60}};
	\draw [{Bar}-{Bar}] ($(n1-Comp.north west)+(0,0.1cm)$)-- ($(n1-Comp.north east)+(0,0.1cm)$) node [midway, above,sloped] {$\overline{w}$};
	\draw [{Bar}-{Bar}] ($(n1-Comp.north west)+(-0.2cm,0)$)-- ($(n1-Comp.south west)+(-0.2cm,0)$) node [midway, left] {$\overline{l}$};
	\draw [{Bar}-{Bar}] (n1-in) ++ (0.15cm,0) coordinate (n1b) -- ($(n1b|-n1-Comp.center)$) node [midway, right] {$\overline{b}$};
	\draw [{Bar}-{Bar}] (n1-Comp.south) ++ (0,-0.15cm) coordinate (n1a) -- ($(n1a-|n1-Comp.east)$) node [midway, left] {$\overline{a}$};

	\pic[right=\xDist cm of n1-Comp.center,inner sep=0pt,anchor=center] (n2) {CompVar={0,-60}};
	\pic[right=\xDist cm of n2-Comp.center,inner sep=0pt,anchor=center] (n3) {CompVar={90,60}};
	\pic[right=\xDist cm of n3-Comp.center,inner sep=0pt,anchor=center] (n4) {CompVar={90,-60}};
	
	\node[below=\yDist cm of n1-Comp.center] (t1) {$(0,0,0)$};
	\node[left=\xDist cm of t1.center,anchor=center] {$(m,n,r)$};
	\node[below=\yDist cm of n2-Comp.center] (t2) {$(0,1,0)$};
	\node[below=\yDist cm of n3-Comp.center] (t3) {$(1,1,1)$};
	\node[below=\yDist cm of n4-Comp.center] (t4) {$(1,0,1)$};

	\pic[below=\yDist cm of t1,inner sep=0pt,anchor=center] (n5) {CompVar={180,60}};
	\pic[right=\xDist cm of n5-Comp.center,inner sep=0pt,anchor=center] (n6) {CompVar={180,-60}};
	\pic[right=\xDist cm of n6-Comp.center,inner sep=0pt,anchor=center] (n7) {CompVar={-90,60}};
	\pic[right=\xDist cm of n7-Comp.center,inner sep=0pt,anchor=center] (n8) {CompVar={-90,-60}};

	\node[below=\yDist cm of n5-Comp.center] (t5) {$(1,1,0)$};
	\node[left=\xDist cm of t5.center,anchor=center] {$(m,n,r)$};
	\node[below=\yDist cm of n6-Comp.center] (t6) {$(1,0,0)$};
	\node[below=\yDist cm of n7-Comp.center] (t7) {$(0,0,1)$};
	\node[below=\yDist cm of n8-Comp.center] (t8) {$(0,1,1)$};

%
	
\end{tikzpicture}  	              
	\caption{Overview of all possible orientations together with the truth table for an example component. To highlight all possible orientations, both connection points (shafts) are marked in different colors. The default orientation is shown in the top left together with the definition of the component dimensions.}
	\label{fig:component_orientation}
\end{figure}

\nobreak\vspace{-5pt}
\begingroup
\allowdisplaybreaks
\begin{small}
\begin{subequations}
	\begin{alignat}{3}\label{eq:shift_CP}
	a_{\lambda,k,i} \geq&&  \overline{a}_{\lambda,k,i}&\cdot (1-2\cdot m_{\lambda,k}) &&-M\cdot r_{\lambda,k} ,\\
	a_{\lambda,k,i} \leq&&  \overline{a}_{\lambda,k,i}&\cdot (1-2\cdot m_{\lambda,k}) &&+M\cdot r_{\lambda,k} ,\\
	a_{\lambda,k,i} \geq&&  \overline{b}_{\lambda,k,i}&\cdot (1-2\cdot n_{\lambda,k}) &&-M\cdot (1-r_{\lambda,k}) ,\\
	a_{\lambda,k,i} \leq&&  \overline{b}_{\lambda,k,i}&\cdot (1-2\cdot n_{\lambda,k}) &&+M\cdot (1-r_{\lambda,k}) ,\\
	\ifextendedversion
	b_{\lambda,k,i} \geq&&  \overline{b}_{\lambda,k,i}&\cdot (1-2\cdot n_{\lambda,k}) &&-M\cdot r_{\lambda,k} ,\\
	b_{\lambda,k,i} \leq&&  \overline{b}_{\lambda,k,i}&\cdot (1-2\cdot n_{\lambda,k}) &&+M\cdot r_{\lambda,k} , \\
	b_{\lambda,k,i} \geq&&  -\overline{a}_{\lambda,k,i}&\cdot (1-2\cdot m_{\lambda,k}) &&-M\cdot (1-r_{\lambda,k}) , \\
	b_{\lambda,k,i} \leq&&  -\overline{a}_{\lambda,k,i}&\cdot (1-2\cdot m_{\lambda,k}) &&+M\cdot (1-r_{\lambda,k}) ,
	\fi
\end{alignat}
\end{subequations}
\end{small}%
\endgroup
where the constraints hold for all ports on components, i.e., for all $ c_{\lambda,k,i} \in C_{\lambda,k}: v_{\lambda,k} \in \mathcal{C}_\lambda$, for all $k \in \mathcal{K}_\lambda$, for all $\lambda \in \{0,\ldots,N_\mathrm{levels}\}$. We apply a big-M formulation~\cite{RichardsHow2005} to appropriately shift the connection points along with the component rotation. Note that these constraints are applied to each port of the component, but with the binary variables applied to the component. Fig.~\ref{fig:component_orientation} shows the resulting eight distinct component orientations, together with a visual representation of the component dimensions. 

\subsection{Interference Constraints}
%
Interference constraints are an important part of any placement or layout problem, since they prevent overlapping between elements in order to obtain feasible solutions. Since we model all elements as rectangles, we can prevent interference by ensuring that every element is either placed above, below, left or right of each other individual element. Although we need constraints between all elements instead of only connected elements, we can identify some properties that allow us to decrease the number of interference constraints. 
First, we only need to prevent interference between elements on the same system level. Since lower-level elements are always contained within a higher-level subsystem, preventing interference between subsystems $s_{\lambda,i}$ and $s_{\lambda,j}$ will inherently prevent interference between $s_{\lambda,i}$ and any element within $s_{\lambda,j}$. 
Second, when there exist multiple subsystems on the same system level, we only need to prevent interference between elements within each subsystem and not across the different subsystems, following the same reasoning. 
As a result, we need to consider $\sum_{i=1}^{|v_{\lambda-1,k}|-1} i$ element pairs in total per system per system level $\lambda$, where $|v_{\lambda-1,k}|$ is the cardinality or the number of elements in the upper level system $v_{\lambda-1,k}$. 
We can now formulate the interference constraints, inspired by the floor-planning problem~\cite{BoydVandenberghe2004}, using a big-M formulation
\par\nobreak\vspace{-5pt}
\begingroup
\allowdisplaybreaks
\begin{small}
\begin{subequations}
\begin{alignat}{4}\label{eq:interference}
	y_{\lambda,j} &- \frac{l_{\lambda,j}}{2} &&\geq y_{\lambda,i} +\frac{l_{\lambda,i}}{2} &&- M\cdot(p_{\lambda,k,z} + q_{\lambda,k,z}),		 \\
	y_{\lambda,j} &+ \frac{l_{\lambda,j}}{2} &&\leq y_{\lambda,i} -\frac{l_{\lambda,i}}{2} &&+ M\cdot(1+p_{\lambda,k,z} - q_{\lambda,k,z}),		 \\	
	x_{\lambda,j} &- \frac{w_{\lambda,j}}{2} &&\geq x_{\lambda,i} +\frac{w_{\lambda,i}}{2} &&- M\cdot(1-p_{\lambda,k,z} + q_{\lambda,k,z}),		 \\
	x_{\lambda,j} &+ \frac{w_{\lambda,j}}{2} &&\leq x_{\lambda,i} -\frac{w_{\lambda,i}}{2} &&+ M\cdot(2-p_{\lambda,k,z} - q_{\lambda,k,z}),	
\end{alignat}
\end{subequations}
\end{small}%
\endgroup
where these constraints should hold for all element pairs in a system, i.e., for all $v_{\lambda,i}, v_{\lambda,j} \in v_{\lambda-1,k} \cup V_\mathrm{0}: i\neq j$, for all $ v_{\lambda-1,k} \in \mathcal{S}_{\lambda-1}$, for all $\lambda \in \{1,\ldots,N_\mathrm{levels}\}$, and where $p_{\lambda,k,z}$ and $q_{\lambda,k,z}$ are binary variables used for the relative placement of each element pair $(v_{{\lambda},i},v_{{\lambda},j})$, with $z \in \{1,\ldots,|v_{\lambda-1,k}|-1\}$ being an identifier for the element pair. By including the relative placement as optimization variables, we can derive alignment constraints as a function of the relative placement, which will be explained in the subsequent section.


\subsection{Alignment Constraints}\label{sec:alignment}

A vehicle powertrain consists of many mechanical and electrical elements. Whilst electrical elements are usually connected through cables, mechanical elements are connected through non-flexible shafts, which requires the connection points to be axially aligned to minimize the friction losses~\cite{SchultzeLienkamp2016}. Therefore, we enforce all mechanical connections to be connected through a straight shaft. 
Formally, this means that the ports comprising the mechanical connection should have either the same x- or y-coordinate and that the ports should be oriented towards each other. Yet both these constraints depend on the relative placement of the element pair.

For components, we constrain the orientation and the location of the port as a function of the relative placement binary variables. To illustrate the dependency on the relative placement, suppose we take the component from Fig.~\ref{fig:component_orientation} in the default orientation and want to connect a component above it to the red shaft. Then we have to rotate the component such that the shaft is at the upper edge, resulting in a counterclockwise rotation of \unit[90]{degrees}, which is equivalent to $(m,r)=(1,1)$. However, as can be imagined, the required shift in orientation also depends on how the component is defined in the library. If we were to connect the blue shaft instead of the red shaft in the previous example, we would have to rotate the component clockwise as opposed to counterclockwise. As a result, we create a truth table as a function of both the edge at which the port is located in the default orientation and the relative placement, which is shown in Table~\ref{tab:truth_table}. From this truth table, we can obtain the logical functions that will be implemented as linear constraints for each orientation variable and connection point edge. However, this truth table is only valid for the component with the outgoing connection, further referred to as the \emph{ego} component. For the other component, referred to as the \emph{connecting} component, we have to swap the \emph{Top} and \emph{Bottom}, and \emph{Right} and \emph{Left} columns, since changing the perspective is equivalent to rotating the component by \unit[180]{degrees}. Table~\ref{tab:truth_functions} shows the logical functions that can be derived from Table~\ref{tab:truth_table} for each orientation variable for the ego component. 
To implement these logical functions as constraints for every edge, we can create different cases, since the edge of the connection point is always known a priori for components. 
As an example, we show the implementation of the orientation constraint for the right edge as
\par\nobreak\vspace{-5pt}
\begingroup
\allowdisplaybreaks
\begin{small}
	\begin{subequations}
 \begin{alignat}{2}\label{eq:Align_orientation}
 	m_{\lambda,k} &\geq&  p_{\lambda,t,z} &- q_{\lambda,t,z}, \\
 	m_{\lambda,k} &\geq& -p_{\lambda,t,z} &+ q_{\lambda,t,z}, \\
 	m_{\lambda,k} &\leq&\  2-p_{\lambda,t,z} &- q_{\lambda,t,z}, \\
 	m_{\lambda,k} &\leq& p_{\lambda,t,z} &+ q_{\lambda,t,z}, \\
 	r_{\lambda,k} &=&\  1-p_{\lambda,t,z}&,
 \end{alignat}
\end{subequations}
\end{small}%
\endgroup
which hold for all outgoing mechanical connections of components, i.e., for all $(c_{\lambda,k,i}, c_{\lambda,l,j})  \in \mathcal{M}_{\lambda}: v_{{\lambda},k} \in \mathcal{C}_\lambda$, for all $\lambda \in \{0,\ldots,N_\mathrm{levels}\}$, with $t$ being an identifier for the (sub)system that includes the connection. 
Note that there is no constraint on $n_{\lambda,k}$, since it is a free variable. The other constraints can be implemented in a similar linear formulation, but since their formulation is mostly trivial, we do not explicitly show them.

\begin{table}[]
	\caption{Truth table showing the required values for the orientation variables $(m,n,r)$ as a function of the relative placement and the edge of the connection point. The $\sim$ indicates that the variable is free. }
	\label{tab:truth_table}
	\centering
	\begin{tabular}{|c|c|c|c|c|}
		\hline
		\textbf{}&\multicolumn{4}{c|}{\textbf{Connection point edge}} \\
		\cline{2-5} 
		\textbf{$(p,q)$} & \textbf{\textit{Top}}& \textbf{\textit{Bottom}}& \textbf{\textit{Right}}&\textbf{\textit{Left}}\\
		\hline
		(0,0) & ($\sim$,0,0) & ($\sim$,1,0) & (1,$\sim$,1) & (0,$\sim$,1) \\
		(0,1) & ($\sim$,1,0) & ($\sim$,0,0) & (0,$\sim$,1) & (1,$\sim$,1) \\
		(1,0) & ($\sim$,0,1) & ($\sim$,1,1) & (0,$\sim$,0) & (1,$\sim$,0) \\
		(1,1) & ($\sim$,1,1) & ($\sim$,0,1) & (1,$\sim$,0) & (1,$\sim$,0) \\
		\hline
	\end{tabular}
\end{table}

\begin{table}[]
	\caption{Logical functions for each orientation binary variable as a function of the edge of the connection point from the ego component perspective. }
	\label{tab:truth_functions}
	\centering
	\begin{tabular}{|l|l|l|l|l|}
		\hline
		\textbf{Orientation}&\multicolumn{4}{c|}{\textbf{Connection point edge}} \\
		\cline{2-5} 
		\textbf{Variable} & \textbf{\textit{Top}}& \textbf{\textit{Bottom}}& \textbf{\textit{Right}}&\textbf{\textit{Left}}\\
		\hline
		$ m $ & Free & Free & $p \odot q $ & $p \oplus q$ \\
		$ n $ & $ q $ & $ 1-q $ & Free & Free \\
		$ r $ & $ p $ & $ p $ & $ 1-p $ & $1-p$\\
		\hline
	\end{tabular}
\end{table}
 
In case of subsystems, there is no need for a shift in orientation, since the dimensions are variable. Whereas the connection point location is free for electrical subsystems, we constrain it to be on the appropriate edge for mechanical subsystems. This is done for the ego perspective by using a big-M formulation according to 
\par\nobreak\vspace{-5pt}
\begingroup
\allowdisplaybreaks
\begin{small}
\begin{subequations}
	\begin{alignat}{2}
		\label{eq:Align_CP_edge1}
	b_{\lambda,k,i} &\geq  \frac{l_{\lambda,k}}{2} &&- M\cdot (p_{\lambda,t,z}+q_{\lambda,t,z}), \\
	b_{\lambda,k,i} &\leq  -\frac{l_{\lambda,k}}{2} &&+ M\cdot (1+p_{\lambda,t,z}-q_{\lambda,t,z}) ,\\
	a_{\lambda,k,i} &\geq  \frac{w_{\lambda,k}}{2} &&- M\cdot (1-p_{\lambda,t,z}+q_{\lambda,t,z}), \\
	a_{\lambda,k,i} &\leq  -\frac{w_{\lambda,k}}{2} &&+ M\cdot (2-p_{\lambda,t,z}-q_{\lambda,t,z}), \label{eq:Align_CP_edge4}
\end{alignat}   
\end{subequations}
\end{small}%
\endgroup
which hold for all outgoing mechanical connections of a subsystem, i.e., for all $(c_{\lambda,k,i}, c_{\lambda,l,j})  \in \mathcal{M}_{\lambda}: v_{{\lambda},k} \in \mathcal{S}_\lambda$, for all $\lambda \in \{0,\ldots,N_\mathrm{levels}\}$.
These constraints will enforce the connection point to be on either of the four edges when combined with~\eqref{eq:CP_in_box1} and~\eqref{eq:CP_in_box2}. For connecting subsystems with an incoming connection, we have to swap the top and bottom, and right and left edges in the constraints.

Finally, we have to constrain either the x- or y-coordinate of the two ports to be equal, both for subsystems and components. Since this constraint is again of a combinatorial nature, we formulate it through a big-M formulation as
\par\nobreak\vspace{-5pt}
\begingroup
\allowdisplaybreaks
\begin{small}
\begin{subequations}
\begin{alignat}{2}
	\label{eq:Align_CP_coord1}
	y_{\lambda,k} + b_{\lambda,k,i} &\geq y_{\lambda,l} + b_{\lambda,l,j} &&- M\cdot (1-p_{\lambda,t,z}), \\
	y_{\lambda,k} + b_{\lambda,k,i} &\leq y_{\lambda,l} + b_{\lambda,l,j} &&+ M\cdot (1-p_{\lambda,t,z}), \\
	x_{\lambda,k} + a_{\lambda,k,i} &\geq x_{\lambda,l} + a_{\lambda,l,j} &&- M\cdot (p_{\lambda,t,z}), \\
	x_{\lambda,k} + a_{\lambda,k,i} &\leq x_{\lambda,l} + a_{\lambda,l,j} &&+ M\cdot (p_{\lambda,t,z}),\label{eq:Align_CP_coord4}
\end{alignat} 
\end{subequations} 
\end{small}%
\endgroup
which holds for all mechanical connections, i.e., for all $(c_{\lambda,k,i}, c_{\lambda,l,j})  \in \mathcal{M}_{\lambda}$, for all $\lambda \in \{0,\ldots,N_\mathrm{levels}\}$, where the first two constraints enforce the y-coordinates of the ports to be equal and the last two constraints enforce the x-coordinates to be equal. Here, we only need the relative placement variable $p_{\lambda,t,z}$, since it is sufficient to know whether we have a vertical or horizontal connection. In some cases, it is desired to have a direct connection, i.e., a connection where the ports are directly connected without an intermediate shaft. For such cases, we can directly constrain both the x- and y-coordinates of the ports to be equal, without the need of any relative placement variable.

When we have a mechanical subsystem, we should not only ensure that the external ports align, but also that the internal elements connected to the external port align in terms of orientation and location. Since the internal element is contained within the subsystem, it should always have its port on the same edge as the subsystem. Yet, the required constraints depend on whether the internal element is a component or subsystem and whether the subsystem is the ego element or connecting element.
If the subsystem is the ego element and if the internal element is a component, we can constrain the orientation of the component according to Table~\ref{tab:truth_functions}, using the placement variables of the parent subsystem. 
Conversely, if the internal element is another subsystem, we have to constrain the port edge according to~\eqref{eq:Align_CP_edge1}-\eqref{eq:Align_CP_edge4}, using the port location and dimension variables of the internal subsystem.
If the subsystem is the connecting element and if the internal element is a component, the \emph{Top} and \emph{Bottom}, and \emph{Right} and \emph{Left} columns in Table~\ref{tab:truth_functions} should be swapped. 
Accordingly, if the internal element is a subsystem, we have to swap the constraints in~\eqref{eq:Align_CP_edge1}-\eqref{eq:Align_CP_edge4} to obtain
\par\nobreak\vspace{-5pt}
\begingroup
\allowdisplaybreaks
\begin{small}
\begin{subequations}
\begin{alignat}{2}
	\label{eq:child_conn_edge1}
	b_{\lambda+1,k,j} &\leq  -\frac{l_{\lambda+1,k}}{2} &&+ M\cdot (p_{\lambda,t,z}+q_{\lambda,t,z}), \\
	b_{\lambda+1,k,j} &\geq  \frac{l_{\lambda+1,k}}{2} &&- M\cdot (1+p_{\lambda,t,z}-q_{\lambda,t,z}), \\
	a_{\lambda+1,k,j} &\leq  -\frac{w_{\lambda+1,k}}{2} &&+ M\cdot (1-p_{\lambda,t,z}+q_{\lambda,t,z}), \\
	a_{\lambda+1,k,j} &\geq  \frac{w_{\lambda+1,k}}{2} &&- M\cdot (2-p_{\lambda,t,z}-q_{\lambda,t,z}), \label{eq:child_conn_edge4}
\end{alignat}
\end{subequations}
\end{small}%
\endgroup
which should hold for all internal mechanical subsystems connected to the upper level subsystem, i.e., for all $(c_{\lambda,l,i},c_{{\lambda+1},k,j}) \in \mathcal{M}_{\lambda,\lambda+1}: v_{\lambda+1,k}\in v_{\lambda,l} \land  v_{\lambda+1,k} \in \mathcal{S}_{\lambda+1}$ , for all $v_{\lambda,l} \in \mathcal{S}_\lambda$, for all $\lambda \in \{0,\ldots,N_\mathrm{levels}-1\}$. 

\begin{figure}[!t]
	\centering
	\tikzstyle{Comp} = [rectangle, minimum width=0.5cm, minimum height=1cm,text centered, draw=black, line width=0.25mm,fill=gray]

\tikzset{
	pics/CompVar/.style args = {#1,#2}{
		code ={
			\node[rectangle, minimum width=0.3cm, minimum height=1cm,text centered, draw=black, line width=0.25mm,fill=gray,rotate=#1] (-Comp) {};
			\draw[line width=0.7mm,red] (-Comp.#2) --++ (#1:0.15) node (-in) {};
			\draw[line width=0.7mm,blue] (-Comp.#2-180) --++ (#1:-0.15) node (-out) {};
			\filldraw[black] (-Comp.center) circle (1pt); 
		}
	}
}

\begin{tikzpicture}

	\tikzstyle{every node}=[font=\footnotesize] 
	
	\definecolor{Fcol}{rgb}{1,0.00,0.00} 
	\definecolor{vcol}{rgb}{0,0.50,0.00} 
	
	\coordinate (Origin) at (0,0);
	\pgfmathsetmacro{\xDist}{1.5}
	\pgfmathsetmacro{\yDist}{0.7}

	\pic[inner sep=0pt,anchor=center] (n1) at (0,0)  {CompVar={0,60}} node[yshift=-0.2cm,label={$v_{2,1}$}] at (n1-Comp.north) {};
	\pic[right=0.15 cm of n1-in.center,inner sep=0pt,anchor=60+180] (n2) {CompVar={0,60}} node[yshift=0.2cm,label={below:$v_{2,2}$}] at (n2-Comp.south) {};
	\node [fit=(n1-out)(n2-in)(n1-Comp)(n2-Comp), draw, thick, inner sep=0pt,label={above:Ego subsystem},align=center,anchor=center] (sys) {} node[above right] at (sys.south east) {$v_{1,1}$};
		
	\pic[right=1.7 cm of n2-in.south west,inner sep=0pt,anchor=-60+180] (n3) {CompVar={0,-60}} node[yshift=-0.1cm,label={Connecting component}] at (n3-Comp.north) {} node[label={below:$v_{1,2}$} ] at (n3-Comp.south) {} ;
	\draw[ultra thick] (n2-in.south west)--(n3-out.north) node [midway,fill=none,below,font=\footnotesize] {}; 
	\filldraw[red,inner sep=0pt,anchor=center,align=center] (n2-in.south-|sys.east) circle (1.5pt); 
	\filldraw[blue,inner sep=0pt,anchor=center,align=center] (n1-out.center-|sys.west) circle (1.5pt); 
	
	\node[below right=-1.2cm and 1cm of n3-Comp] (tab1) {%
		\begin{tabular}{c|c}
			Variables & Values\\
			\hline 
			$(p_{1,1,1},q_{1,1,1})$ & $(1,0)$\\
			$(m_{2,2},n_{2,2},r_{2,2})$ & $(0,0,0)$\\
			$a_{1,1,1}$ & $\frac{{w}_{1,1}}{2}$\\
			$a_{2,2,1}$ & $\frac{\overline{w}_{2,2}}{2}$\\
		\end{tabular}
	};
	
%
%
%
%
%
	
	%
	
\end{tikzpicture}  	              
	\caption{Example of a mechanical subsystem connected to a component. The ports of the subsystem align with the connection points of the internal components. In addition, the values are shown of the most relevant placement and orientation variables for the connection $(v_{1,1},v_{1,2})$ and component $v_{2,2}$. }
	\label{fig:subsystem_internal}
\end{figure}
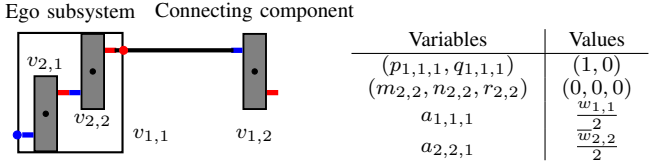

Irrespective of the type of internal element, we can constrain the location of the port using~\eqref{eq:Align_CP_coord1}-\eqref{eq:Align_CP_coord4} with the left-hand sides replaced by the location of the port of the internal element. This applies to both the ego perspective as well as the connecting perspective, since we only need to distinguish between horizontal or vertical connections. To summarize, Fig.~\ref{fig:subsystem_internal} shows an example of a mechanical connection between a subsystem with internal components and an external component together with some of the placement and orientation variables that satisfy the aforementioned constraints.

\subsection{Optimization Problem}
This section presents the optimal placement problem for the electric vehicle powertrain elements.  Given a predefined library of elements and topology, we formulate the design problem as a mixed-integer convex quadratic program, whereby the objective function can be of any (multi-objective) quadratic or linear formulation. Examples of an objective function specifically for powertrain design could be to minimize the connection lengths and subsystem dimensions, maximize passenger space, or to obtain a certain weight distribution. In the remainder of the paper, our objective is to minimize the unweighted sum of connection lengths and subsystem dimensions, given by
	\par\nobreak\vspace{-5pt}
	\begingroup
	\allowdisplaybreaks
	\begin{small}
		\begin{alignat}{2}
		\begin{split}
	&J_\mathrm{con} = \sum_{\lambda=0}^{N_\mathrm{levels}}\!\!\!\! \sum_{(c_{\lambda,k,i}, c_{\lambda,l,j}) \in E_\lambda}\begin{Vmatrix}
	x_{\lambda,k}+a_{\lambda,k,i}-x_{\lambda,l}-a_{\lambda,l,j}\\
	y_{\lambda,k}+b_{\lambda,k,i}-y_{\lambda,l}-b_{\lambda,l,j}
\end{Vmatrix}_2 + \\
 &\!\sum_{\lambda=0}^{N_\mathrm{levels-1}}\!\!\!\!\!\! \sum_{(c_{\lambda,k,i}, c_{\lambda+1,l,j}) \in E_{\lambda,\lambda+1}}\begin{Vmatrix}
	x_{\lambda,k}+a_{\lambda,k,i}-x_{\lambda+1,l}-a_{\lambda+1,l,j}\\
	y_{\lambda,k}+b_{\lambda,k,i}-y_{\lambda+1,l}-b_{\lambda+1,l,j}
\end{Vmatrix}_2\!, 
\end{split} \\
&J_\mathrm{dim} = \sum_{\lambda=0}^{N_\mathrm{levels}} \sum_{i=0}^{|\mathcal{S}_\lambda|} (w_{\lambda,i}+l_{\lambda,i}).
\end{alignat}
\end{small}%
\endgroup
Using the design variables $d=(x,y,w,h,a,b) $ and the binary inputs $u=(p,q,m,n,r)$, we frame the optimization problem as follows:
\begin{prob}[Optimal element placement]\label{prob:placement}
	The optimal element placement is the solution of
	\par\nobreak\vspace{-5pt}
	\begingroup
	\allowdisplaybreaks
	\begin{small}
	\begin{equation*}
				 \begin{aligned}
			&\min &&J_\mathrm{con} + J_\mathrm{dim}, \\
			&\textnormal{s.t. }  &&\eqref{eq:El_in_box} -\eqref{eq:child_conn_edge4}, \\ &&&\textnormal{Other alignment constraints according to Table~\ref{tab:truth_functions}} .\\
		\end{aligned}
	\end{equation*}
	\end{small}%
	\endgroup
\end{prob}
\noindent
Since Problem 1 can be solved with mixed-integer convex quadratic programming solvers, we can guarantee global optimality~\cite{Lee2012}.
\section{Results } \label{Results}



This section presents numerical results for the two-dimensional placement problem. We consider a design space bounded by the vehicle body from the top view, where we minimize connection lengths and subsystem dimensions. As a case study, we consider the powertrain topology shown in Fig.~\ref{fig:topology}, which describes a four-wheel driven electric vehicle. There are three subsystems present in total, being the battery pack and two transmissions. The battery pack consists of 24 modules connected in series, whereas the transmissions consist of two gear pairs connected in series. To improve the computation time, we group the battery modules in four blocks. While this comes at the cost of a slight reduction in design freedom, it has a major impact on the search space to be explored by the solver. 

We parse the optimal placement problem with YALMIP~\cite{Loefberg2004} and solve it using Gurobi~\cite{GurobiOptimization2021}. We perform the numerical optimization on an Intel Core i7-4710MQ 2.5GHz processor and 8GB of RAM. Thereby, the total computation time was about \unit[2]{min} for this specific use case.  

\begin{figure}[!t]
	\centering
	\input{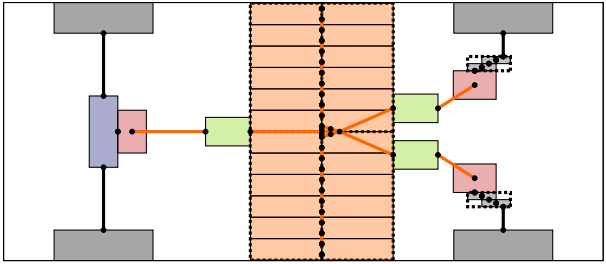}                
	\caption{Optimal element placement and subsystem (dashed) dimensions for the reference powertrain topology. All mechanical connections are axially aligned and connected through shafts (black), whereas the electrical connections (orange) are free.}
	\label{fig:solution}
\end{figure}

Fig.~\ref{fig:solution} shows the optimal element placement for the reference topology. First, we observe that all mechanical connection ports (shown in black) are aligned, whereas the electrical connections (in orange) remained free. Second, we enforced the mechanical port of the \gls{acr:em} to be directly connected, resulting in the connecting ports to have the same location. 
Third, we notice that the battery pack is as short as possible in the driving direction, since this allows the inverters to be as close as possible to the (single) battery connection point. Furthermore, the battery modules are oriented towards the center of the pack and the battery pack connection port is shifted rearward, both resulting in short (internal) connections.
Finally, we observe a large difference between the typical and the optimal solution in element placement near the rear axle. Whereas the typical solution often places the transmission, \gls{acr:em} and inverter as close as possible to each other, the optimal solution minimizes the connections from a systems perspective. Specifically, it positions the gear pairs to the front, thereby trading an increase in transmission size for a lower overall connection length. This highlights the importance of a holistic design approach and the advantage of jointly optimizing across multiple system levels. 

%





\section{Conclusion} \label{Conclusion}


In this paper, we studied the spatial design problem for vehicle powertrains, explicitly accounting for rotation, interference and alignment constraints. To this end, we devised an optimization framework that can compute the optimal element placement from a systems perspective, thereby contributing to a holistic systems design approach. 
Using our framework, we are able to automatically generate system designs with reasonable computation times, whilst respecting alignment constraints.

This work opens the field for the following possible extensions: First, we want to integrate the method within existing topology generation and powertrain optimization frameworks to obtain a fully automated holistic design approach.
Second, we want to extend the framework to the three-dimensional placement problem and explore different objective functions. 
%

\section*{Acknowledgment}
We thank Dr. I. New, Ir. F. Vehlhaber, Ir. M. Clemente and Ir. F. Paparella for proofreading this paper, and Ir. L. Pedroso for the suggestions on the mathematical notations.
\bibliographystyle{IEEEtran}
\bibliography{main,SML_papers}

\end{document}